\documentclass{article}
\usepackage{spconf,amsmath,amsfonts,amssymb,graphicx,hyperref}
\usepackage{array,booktabs,textcomp,url,cite}
\usepackage{xcolor}

\definecolor{green}{rgb}{0, 0.5, 0}
\definecolor{orange}{rgb}{0.8, 0.6, 0.2}
\definecolor{orange2}{rgb}{1.0, 0.6, 0.2}
\definecolor{red}{rgb}{1.0, 0.0, 0.0}
\definecolor{teal}{rgb}{0.0, 0.4, 0.4}
\definecolor{purple}{rgb}{0.65,0,0.65}
\definecolor{saffron}{rgb}{0.95,0.75,0.2}
\definecolor{turquoise}{rgb}{0.0,0.5,0.5}
\definecolor{black}{rgb}{0.0, 0.0, 0.0}
\definecolor{gray}{rgb}{0.5, 0.5, 0.5}

\newcommand{\yang}[1]{{\color{black}#1}}
\newcommand{\ying}[1]{{\color{black}#1}}
\newcommand{\zhikai}[1]{{\color{black}#1}}

\title{ReVR: Dual-Path Concept Reasoning for Multimodal Fake News Detection}

\name{Zhikai Tan$^{\star}$, Yuzhou Yang$^{\star}$, Qichao Ying$^{\star}$, Pinjie Xu$^{\dagger}$, Sheng Li$^{\star}$, Zhenxing Qian$^{\star}$ and Xinpeng Zhang$^{\star}$}
\address{$^{\star}$ Fudan University, $^{\dagger}$ China University of Mining and Technology - Beijing}

\begin{document}

\maketitle

\begin{abstract}
Vision-language models (VLMs) support multimodal fake news detection (FND) by producing explicit analyses.
Recent methods further improve interpretability by organizing verification knowledge into explicit concepts.
However, two questions remain: how to improve the reliability and applicability of verification concepts, and how to effectively apply reusable concepts to verify unseen news.
We propose \textbf{ReVR}, a dual-path reasoning framework that constructs and applies reusable verification concepts for multimodal fake news detection.
An agentic workflow grounds and consolidates candidate concepts, while statistical profiles characterize their historical behavior.
During inference, a coverage-oriented path aggregates evidence from the complete concept library using a trainable encoder, while a query-focused path prompts a frozen VLM to reason over selected concepts and their observations.
A learned conflict resolver selects between the two predictions when they disagree.
Experiments on fake news benchmarks demonstrate the effectiveness of the method regarding detection performance and generalizability.
\end{abstract}


\section{Introduction}

Multimodal fake news detection aims to determine news veracity by jointly analyzing textual and visual content.
The task involves assessing both the authenticity of each modality and the consistency of their content.
Existing approaches learn multimodal representations through text-image fusion~\cite{wang2018eann,khattar2019mvae} and model cross-modal consistency or ambiguity~\cite{chen2022cafe}.
More fine-grained analysis examines image and text manipulation cues~\cite{shao2023hammer}, local cross-modal correlations~\cite{qiao2025c3n,zhou2023multimodal}, and alignment of entities, events, and temporal information~\cite{guo2025amg}.
However, individual verification cues can vary in relevance and reliability across news items, and inaccurate intermediate assessments may affect the final prediction.
This motivates characterizing their behavior across samples and studying how to apply them effectively to unseen news.

\ying{Although news items differ in topic and presentation, verification often relies on recurring questions about text-image consistency, entity agreement, and factual support.}
\yang{Thus, beyond producing free-form rationales, a growing line of work converts recurring verification cues into explicit intermediate knowledge that can be inspected by human and reused by models~\cite{yang2026roe}.}
\yang{These studies extend feature-based prediction toward explicit reasoning over human-interpretable concepts and their relations, building on concept bottlenecks~\cite{koh2020concept} and learned logical rules~\cite{bhattarai2022explainable}.}
\ying{TELLER~\cite{liu2024teller} uses large language models to evaluate expert-defined fact-checking predicates and learns interpretable logical rules.
MiRAGe~\cite{huang2024miragenews} combines object-level visual concepts and automatically discovered textual concepts with embedding-based classifiers.
PCGR~\cite{yang2026pcgr} discovers interrogative concepts from high-loss multimodal samples, organizes them into an evolving layered graph, and performs hierarchical probabilistic reasoning.
}

\yang{Despite improved interpretability and performance of the literature works, two questions remain:
1) \textit{How can we improve the reliability and applicability of proposed verification concepts?}
Both predefined predicates and automatically discovered concepts may be biased or limited in scope, and VLM-generated analyses may contain hallucinations.}
\yang{As simple filtering, such as deduplication, may leave these issues unresolved,
grounding concepts in source evidence and building historical statistical profiles are important for assessing their reliability and applicability.}
2) \textit{How can we further improve the effectiveness of applying reusable concepts to verify unseen news?}
Assessing many concepts may include weakly relevant cues, whereas detailed reasoning over a selected subset may overlook complementary evidence.
\ying{So beyond modeling concept dependencies, effective verification must balance broad assessment with contextual reasoning,
}

\begin{figure*}[t]
\centering
\includegraphics[width=1.\textwidth]{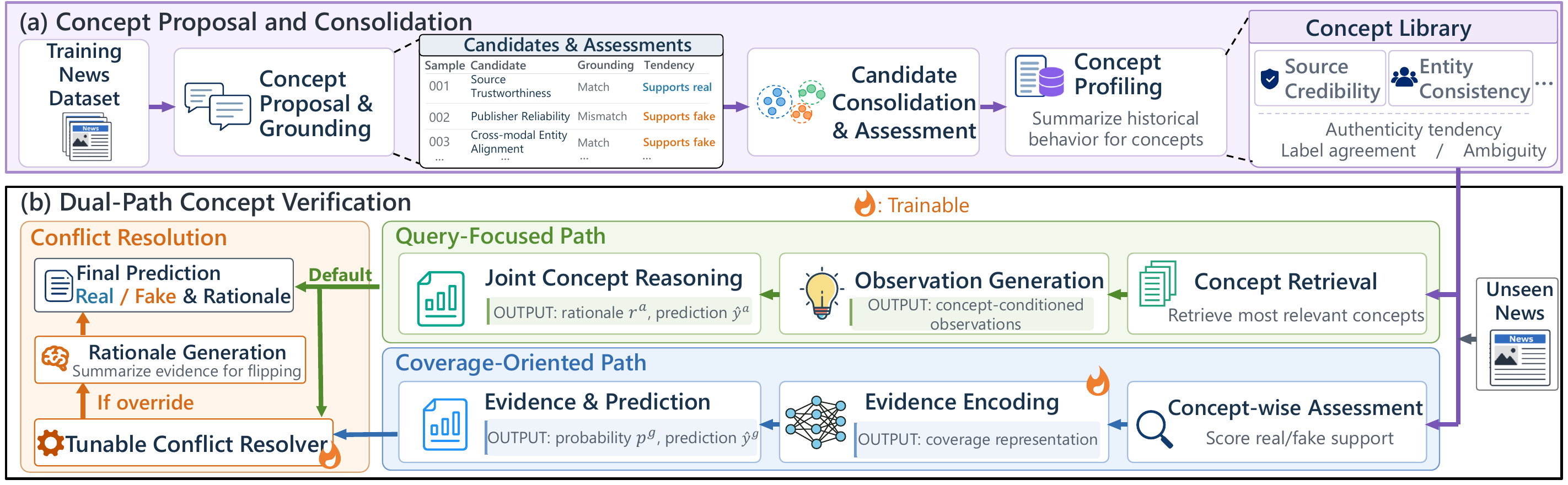}
\caption{
Framework overview of ReVR. (a) We construct a reusable concept library through grounded proposal, iterative consolidation, and historical profiling.
(b) Query-focused and coverage-oriented verification assess unseen news, with learned conflict resolution selecting between disagreeing predictions.
}
\label{fig:framework}
\end{figure*}

We propose \textbf{ReVR}, a dual-path reasoning framework that constructs and applies reusable verification concepts for multimodal fake news detection.
We first build a reusable concept library by prompting a frozen VLM to propose candidates from training news, ground their analyses in source evidence, and consolidate related candidates into canonical concepts.
We check each canonical concept against its associated samples and summarize its historical behavior in a statistical profile.
For unseen news, the query-focused path prompts the same frozen VLM to reason over selected concepts, while the coverage-oriented path uses a trainable encoder and classifier to aggregate evidence from the full library with historical profiles.
A trainable conflict resolver retains the paths' common verdict or selects between them on disagreement.
Experiments on multimodal fake news benchmarks demonstrate the effectiveness of the method regarding detection performance and generalizability.


\section{Proposed Method}
\label{sec:method}

Figure~\ref{fig:framework} presents the two stages of ReVR, i.e., the Concept Proposal and Consolidation stage, and the Dual-path Concept Verification stage.

\subsection{Concept Proposal and Consolidation}
\label{subsec:concept_induction}
\ying{Given training news items $x=(T,I)$ with text $T$, image $I$, and veracity label $y$, we generate and ground raw candidate concepts $\tilde c$.
We consolidate related candidates into $M$ canonical concepts $\{c_m\}_{m=1}^{M}$ and profile their historical behavior to obtain a reusable concept library $\mathcal C$.}


\noindent\textbf{Candidate Concept Proposal \& Grounding.}
A lightweight agentic workflow prompts a frozen VLM to construct concepts in three steps.
1) \textit{Thinking}: For each training item $x$, it proposes distinct raw candidate concepts $\tilde c$.
Each candidate contains a stance-neutral verification angle, a sample-specific
analysis, and an assessment of whether the evidence favors real
news, fake news, or neither.
2) \textit{Inspecting}: It invokes local tools as needed, including text-span inspection,
OCR, and image cropping, to locate supporting text spans or image
regions.
The thinking and inspecting steps repeat until the VLM emits an exit status.
3) \textit{Grounding}: The VLM checks whether the located evidence supports the analysis
and is relevant to the proposed angle, and it can directly reject candidates that it claims unsupported.
The retained candidates' semantic content is used for consolidation,
while their real/fake assessments are retained for profiling.

\noindent\textbf{Concept Consolidation \& Assessment.}
\zhikai{To construct reusable verification angles with consistent
cross-instance applicability, ReVR consolidates the generated candidates in
two iterative stages. 1) \textit{Grouping}: The VLM groups semantically similar retained candidates and removes duplicates. Each group is represented by a canonical concept $c_m$.
2) \textit{Assessing and refining}: The VLM checks each
canonical concept $c_m$ against its group members, and refines and rechecks
concepts with member recall below an empirical threshold of 0.7.
Accordingly, concepts whose cluster center fails the recall requirement after three modifications are released again to the candidate pool and we redo the consolidation. 
After iterative consolidation and assessment, we let VLM assign each retained
group a concise canonical name and verification scope for $c_m$, and the retained canonical concepts are then used for profiling and inference.}

\noindent\textbf{Concept Profiling.}
\zhikai{For each canonical concept $c_m$, we profile
three complementary historical properties. Specifically,
$\boldsymbol{\pi}_m=[b_m,\rho_m,u_m]$. 1) \textit{Authenticity tendency} $b_m$ is the average signed direction, with positive values favoring fake news and negative values favoring real news.
2)\textit{ Label agreement} $\rho_m$, the fraction of applicable assessments matching the ground-truth labels, provides
a reliability prior. 3) \textit{Ambiguity} $u_m$, measured as the frequency of
non-directional assessments, enables uncertainty-aware weighting.}
The complete library is
$\mathcal C=\{(c_m,\boldsymbol{\pi}_m)\}_{m=1}^{M}$
, where
$M$ is the number of retained canonical concepts and $\boldsymbol{\pi}_m$ is the historical profile of $c_m$. Profiles are computed from
training data and fixed during testing.

\subsection{Dual-path Concept Verification}
\ying{Given an unseen news item $x$ and library $\mathcal C$, ReVR obtains predictions through query-focused ($a$) reasoning over selected concepts and coverage-oriented ($g$) evidence encoding across the full library.
We retain their common verdict or select between disagreeing predictions using a learned conflict resolver, and return the final prediction $\widehat y$ with a corresponding rationale for interpretation.}


\noindent\textbf{Query-Focused Concept Reasoning.}
To capture instance-specific evidence, the frozen VLM performs three steps. 1) \textit{Retrieving}: A retrieval prompt sequentially selects $k$ distinct concepts from $\mathcal C$ that are most relevant to $x$. 2) \textit{Observing}: An observation prompt examines the news text and image under each selected concept's verification scope. And 3) \textit{Aggregating}: An aggregation prompt jointly reasons over the resulting concept-observation pairs to produce a natural-language rationale and prediction $\widehat y^a$. This path is fully prompt-driven without tunable parameters.
\begin{figure*}[t]
\centering
\includegraphics[width=1.\textwidth]{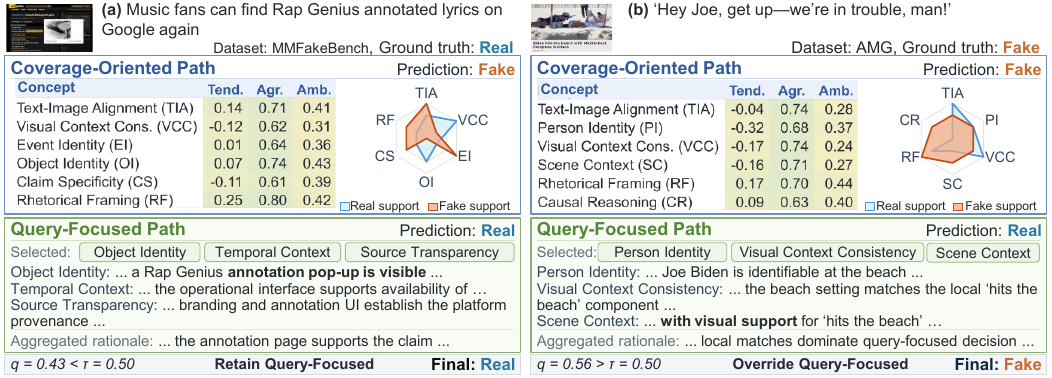} 
\caption{\zhikai{Case studies of ReVR. Tend., Agr., and Amb. denote historical tendency, label agreement, and ambiguity, with positive tendency favoring fake news. Radar plots show VLM-generated support scores for the real (blue, $s_m^R$) and fake (orange, $s_m^F$) hypotheses under each displayed concept. $q$ and $\tau$ denote the override probability and threshold. Conflict-resolution rationales are omitted for space.}}
\label{fig:case_study}
\end{figure*}

\noindent\textbf{\zhikai{Coverage-Oriented Concept Reasoning.}}
\zhikai{To get complementary evidences over all discovered concepts, the coverage-oriented path assesses $x$ under every concept in $\mathcal C$, preserving evidence for both veracity hypotheses~\cite{zheng2025aar,chen2026mmrgv}:}
\begin{equation}
(s_m^R,s_m^F)=\operatorname{CA}(x,c_m).
\label{eq:bidirectional_scores}
\end{equation}
Here, $\operatorname{CA}$ denotes \textit{coverage-oriented assessing}: a fixed-prompt call to the frozen VLM that separately returns real and fake support scores $s_m^R$ and $s_m^F$ for each concept $m=1,\ldots,M$.
Three trainable modules combine these scores with concept semantics and historical profiles:
\begin{equation}
\begin{aligned}
\mathbf h_m^g &= \phi_g(s_m^R,s_m^F,c_m,\boldsymbol{\pi}_m),\\
\mathbf z^g &= \operatorname{Enc}_g\left(\{\mathbf h_m^g\}_{m=1}^{M}\right),\\
p^g &= \operatorname{Cls}_g(\mathbf z^g).
\end{aligned}
\label{eq:global_representation}
\end{equation}
Here, $\phi_g$ is a shared projection forming concept-wise evidence tokens $\mathbf h_m^g$, $\operatorname{Enc}_g$ is a Transformer encoder~\cite{vaswani2017attention} aggregating them into the full-library evidence representation $\mathbf z^g$, and $\operatorname{Cls}_g$ is a classification head producing the fake-news probability $p^g$, from which prediction $\widehat y^g$ is obtained.
These modules are trained against the ground-truth label $y$ using binary cross-entropy:
$
\mathcal L_g=-\mathbb E\!\left[y\log p^g+(1-y)\log(1-p^g)\right],
$
where $\mathbb E$ averages over training samples.

\noindent\textbf{\zhikai{Conflict Resolution with Conditional Reliability Estimation.}}
\zhikai{On disagreement between the two paths, a trainable $\ell_2$-regularized logistic resolver $F_{\mathrm{cr}}$ estimates the probability $q$ that the coverage-oriented prediction is correct:}
\begin{equation}
\zhikai{q=F_{\mathrm{cr}}(p^g,\mathbf z^g,\widehat y^a).}
\label{eq:routing_rule}
\end{equation}
Its continuous inputs are standardized before fitting.
We train the resolver on disagreeing samples with the correctness target $r=\mathbb I[\widehat y^g=y]$ using binary cross-entropy:
$
\mathcal L_{\mathrm{cr}}=-\mathbb E_{\widehat y^g\neq\widehat y^a}\!\left[r\log q+(1-r)\log(1-q)\right].
$
Here, $\mathbb I[\cdot]$ is the indicator function, and the expectation is restricted to disagreeing training samples.
Because the labels are binary, $r=0$ indicates that the query-focused prediction is correct.
We retain the common verdict when the paths agree.
On disagreement, the final verdict $\widehat y$ is the coverage-oriented prediction when $q\geq\tau$ and the query-focused prediction otherwise.
Here, $\tau$ is the override threshold.

\ying{Finally, to improve human-readable interpretability, on a coverage-oriented override, we prompt the frozen VLM with both paths' outputs, concept-wise real/fake support scores, and historical profiles to explain the evidence supporting that verdict.
Otherwise, we return the selected concepts, their observations, and the query-focused rationale.}

\begin{table}[!t]
\caption{Coarse-level accuracy and F1 (\%). Best and second-best results are bold and underlined.}
\label{tab:coarse_mmd_results}
\centering
{
\small
\setlength{\tabcolsep}{1mm}
\begin{tabular*}{\columnwidth}{
    @{\extracolsep{\fill}}
    l
    r r
    r r
    r r
    @{}
}
\toprule
\textbf{Method}
& \multicolumn{2}{c}{\shortstack[c]{\textbf{MMFake}\\\textbf{Bench}}}
& \multicolumn{2}{c}{\textbf{AMG}}
& \multicolumn{2}{c}{\shortstack[c]{\textbf{MiRAGe}\\\textbf{News}}} \\
\cmidrule(lr){2-3}
\cmidrule(lr){4-5}
\cmidrule(l){6-7}
& \textbf{Acc.} & \textbf{F1}
& \textbf{Acc.} & \textbf{F1}
& \textbf{Acc.} & \textbf{F1} \\
\midrule

\multicolumn{7}{@{}l}{\textit{General VLMs}} \\
\addlinespace[1pt]
GPT-5~\cite{openai2025gpt5}
& 58.8 & 57.2
& 59.9 & 57.9
& 56.8 & 54.0 \\
Qwen3.5-Plus~\cite{qwenteam2026qwen35}
& 72.5 & 71.2
& 69.4 & 69.4
& 76.1 & \underline{75.7} \\

\addlinespace[3pt]
\hline
\multicolumn{7}{@{}l}{\textit{Multimodal Detectors}} \\
\addlinespace[1pt]
MPFN~\cite{jing2023mpfn}
& 53.4 & 49.6
& 55.4 & 52.7
& 49.9 & 45.5 \\
BMR~\cite{ying2023bmr}
& 52.3 & 48.8
& 57.3 & 55.4
& 49.5 & 44.8 \\
HAMMER~\cite{shao2023hammer}
& 55.7 & 51.7
& 60.1 & 57.2
& 52.8 & 50.4 \\
C3N~\cite{qiao2025c3n}
& 73.6 & 70.3
& 75.3 & 72.6
& 70.4 & 66.0 \\
MGCA~\cite{guo2025amg}
& 74.1 & 71.3
& 78.2 & 76.8
& 72.3 & 66.6 \\
PCGR~\cite{yang2026pcgr}
& \underline{80.6} & \underline{73.5}
& \textbf{84.3} & \underline{79.8}
& \underline{80.2} & 70.9 \\

\midrule
\textbf{ReVR}
& \textbf{83.0} & \textbf{78.7}
& \textbf{84.3} & \textbf{83.1}
& \textbf{85.8} & \textbf{85.2} \\
\bottomrule
\end{tabular*}
}
\end{table}

\section{Experiments and Results}

\setcounter{topnumber}{3}
\setcounter{totalnumber}{4}

\noindent\textbf{Experimental Setups.}
All VLM calls use Qwen3.5-Plus as backend~\cite{qwenteam2026qwen35}
For each training instance, we generate three candidate concepts and encode them with frozen \texttt{mpnet-v2}~\cite{reimers2019sentencebert,song2020mpnet}.
We apply average-linkage agglomerative clustering over cosine distances with threshold $\delta=0.25$ and retain clusters with no less than an empirical ten instances for each.
The encoder uses a 768-to-32 semantic projection and a two-layer, four-head Transformer 
and is trained with AdamW optimizer with learning rate $10^{-3}$, batch size 64, and 100 epochs.
The query-focused path sequentially retrieves $k=3$ non-repeated concepts.
As for datasets, following the evaluation protocol of many baselines~\cite{yang2026pcgr}, we evaluate on MMFakeBench~\cite{liu2025mmfakebench}, AMG~\cite{guo2025amg}, and MiRAGeNews~\cite{huang2024miragenews}.

\noindent\textbf{Results and Case Studies.}
Table~\ref{tab:coarse_mmd_results} first provides coarse-level accuracy and F1 score of multimodal FND.
\ying{Also, we extend both paths to multi-class prediction on MMFakeBench and AMG, where the conflict resolution stage can also check misalignments between the two paths w.r.t the predicted subtypes.} 
Fig.~\ref{fig:case_study} illustrates the paths' complementary roles. In (a), query-focused reasoning grounds the claim in decisive contextual details despite misleading library-wide evidence. In (b), local visual matches mislead that path, while coverage-oriented reasoning combines broader evidence with concepts' historical behavior. The conflict resolver chooses the correct path in each case.
We see that ReVR consistently outperforms prior methods, supporting the generality of reusable verification concepts across different FND benchmarks.

\begin{table}[!t]
\caption{Fine-grained F1 (\%) on MMFakeBench and AMG}
\label{tab:fine_grained_results}
\centering
{
\small
\setlength{\tabcolsep}{1mm}
\begin{tabular}{@{}lcccc@{}}
\toprule
\textbf{Method}
& \multicolumn{2}{c}{\textbf{MMFakeBench}}
& \multicolumn{2}{c}{\textbf{AMG}} \\
\cmidrule(lr){2-3}
\cmidrule(l){4-5}
& \textbf{Micro-F1} & \textbf{Macro-F1}
& \textbf{Micro-F1} & \textbf{Macro-F1} \\
\midrule
GPT-5~\cite{openai2025gpt5}
& 55.3 & 54.9
& 60.7 & 55.5 \\
Qwen3.5-Plus~\cite{qwenteam2026qwen35}
& 57.6 & 58.0
& 51.0 & 57.1 \\
MGCA~\cite{guo2025amg}
& 60.3 & 55.1
& 72.8 & 57.2 \\
PCGR~\cite{yang2026pcgr}
& 68.6 & 56.9
& \textbf{75.6} & 59.9 \\
\midrule
\textbf{ReVR}
& \textbf{68.8} & \textbf{62.3}
& 73.2 & \textbf{62.1} \\
\bottomrule
\end{tabular}
}
\end{table}

\noindent\textbf{Cross-dataset Generalization.}
We train each method on one source dataset and test it on the other two without target-dataset adaptation. ReVR ranks first in every direction and raises average Macro-F1 from 49.0 to 68.5, showing that we can successfully mine reusable verification concepts with generalizability beyond their source dataset.

\begin{table}[!t]
\caption{Cross-dataset Macro-F1 (\%). M, A, and N denote MMFakeBench, AMG, and MiRAGeNews. }
\label{tab:cross_dataset}
\centering
{
\small
\setlength{\tabcolsep}{0.8mm}
\begin{tabular}{@{}lccccccc@{}}
\toprule
\textbf{Method}
& \textbf{M$\to$A} & \textbf{M$\to$N}
& \textbf{A$\to$M} & \textbf{A$\to$N}
& \textbf{N$\to$M} & \textbf{N$\to$A}
& \textbf{Avg.} \\
\midrule
MPFN~\cite{jing2023mpfn}
& 32.5 & 56.9 & 27.2 & 41.0 & 53.3 & 41.5 & 42.1 \\
BMR~\cite{ying2023bmr}
& 40.7 & 46.1 & 47.4 & 61.5 & 48.1 & 44.6 & 48.1 \\
HAMMER~\cite{shao2023hammer}
& 43.0 & 68.6 & 42.5 & 51.9 & 49.0 & 39.1 & 49.0 \\
C3N~\cite{qiao2025c3n}
& 37.9 & 55.1 & 26.9 & 32.5 & 56.8 & 42.5 & 42.0 \\
PCGR~\cite{yang2026pcgr}
& 35.8 & 51.0 & 43.0 & 55.1 & 56.2 & 44.0 & 47.5 \\
\midrule
\textbf{ReVR}
& \textbf{73.5} & \textbf{72.1} & \textbf{53.8} & \textbf{70.4}
& \textbf{76.4} & \textbf{64.9} & \textbf{68.5} \\
\bottomrule
\end{tabular}
}
\end{table}

\begin{table}[!t]
\caption{Ablation results on the three benchmarks (\%)}
\label{tab:ablation}
\centering
{
\small
\setlength{\tabcolsep}{1mm}
\begin{tabular}{lcccccc}
\toprule
& \multicolumn{2}{c}{\shortstack{MMFake\\Bench}}
& \multicolumn{2}{c}{AMG}
& \multicolumn{2}{c}{\shortstack{MiRAGe\\News}} \\
\cmidrule(lr){2-3}
\cmidrule(lr){4-5}
\cmidrule(lr){6-7}
Variant
& Acc. & F1
& Acc. & F1
& Acc. & F1 \\
\midrule
w/o Coverage Path
& 79.4 & 74.2 & 76.7 & 74.1 & 84.7 & 84.6 \\
w/o Query Path
& 78.2 & 75.5 & 80.7 & 79.8 & 76.4 & 75.8 \\
w/o Concept Profiling
& 79.3 & 74.3 & 80.2 & 79.6 & 81.9 & 81.5 \\
w/o Conflict Resolver
& 80.3 & 76.2 & 82.4 & 81.2 & 83.4 & 82.8 \\
\midrule
\textbf{ReVR}
& \textbf{83.0} & \textbf{78.7}
& \textbf{84.3} & \textbf{83.1}
& \textbf{85.8} & \textbf{85.2} \\
\bottomrule
\end{tabular}
}
\end{table}

\noindent\textbf{Ablation Studies.}
As shown in Table~\ref{tab:ablation}, we construct four variants
while keeping all remaining settings unchanged. Here, ``w/o concept profiling'' means letting the global path freely check all concepts similar to query-focused path, and ``w/0 conflict resolver'' means replacing the tunable resolver with a reasoning VLM for judgement.
We see that every
ablation underperforms the complete model, confirming that the four components
play distinct rather than redundant roles. 
The library-wide encoding is particularly useful for
diverse manipulation patterns, 
and the consistent degradation without profiles shows that
historical behavior helps contextualize current evidence.

\section{Conclusion}
We propose ReVR, which consolidates VLM-generated analyses into reusable verification concepts.
Complementary query-focused and coverage-oriented paths apply these concepts to unseen news, with a learned conflict resolver selecting between their predictions when they disagree.
Extensive experiments validate the effectiveness of the method.

\bibliographystyle{IEEEbib}
\bibliography{main_icassp}

\end{document}